# S-Boxهای پویای وابسته به کلید سبک وزن مبتنی بر خم ابربیضوی برای دستگاه‌های اینترنت اشیا

پروانه اصغری [1*]، سید حمید حاج سید جوادی [2]

[1] دانشکده علوم، گروه ریاضی و علوم کامپیوتر، دانشگاه شاهد، تهران، ایران
[2] گروه مهندسی کامپیوتر، واحد تهران مرکزی، دانشگاه آزاد اسلامی، تهران، ایران

## چکیده

**موضوع امنیت یکی از اصلی ترین مباحث در محیط اینترنت اشیا است. با توجه به اهمیت نقش موثر روش های رمزنگاری بلوک در ایجاد امنیت در اینگونه سیستم‌ها، تولید S-Box از اهمیت ویژه‌ای در رمزنگاری برخوردار است. با توجه به محدودیت ظرفیت منابع درگره های اینترنت اشیا ، تولیدS-Box سبک وزن یک چالش مهم است. در این مقاله یک روش تغییر در S-Boxهای رمزنگاری متقارن ایستا وابسته به کلید و تولید آنها به شکل پویا با استفاده از خم ابربیضوی[1] ارائه می‌شود. S-Box پیشنهادی با استفاده از معیارهای عملکردی از جمله دو سوئی بودن[2]، غیرخطی بودن[3]، اثر فروپاشی بهمنی[4] و درجه جبری[5] ارزیابی می‌شود. نتایج ارزیابی تائید می‌کند که الگوریتم تولید S-Box ارائه شده یک روش موثر برای تولید S-Boxهای سبک وزن و قوی رمزنگاری است.**

## ۱- مقدمه

با توجه به نفوذ روز افزون استفاده از اینترنت اشیا (IoT) در بیشتر جنبه‌های زندگی بشر امروزی، ابعاد مختلف استفاده از این بستر، طی سال‌های اخیر مورد توجه جامعه پژوهشگران قرار گرفته است. هنگام استفاده از اینترنت اشیا بخصوص در کاربردهای حساس و بحرانی، بحث امنیت و رمزنگاری اطلاعات به یک ضرورت حیاتی تبدیل می‌شود. دلایل عمده توسعه روش‌های جدید رمزنگاری سبک وزن در محیط اینترنت اشیا عبارت است از الف) محدودیتهای منابع دستگاه‌های اینترنت اشیا به لحاظ حافظه و توانایی پردازش و ب) بهره وری از ارتباطات انتها به انتها[6] که منجر به استفاده از روش‌های رمزنگاری متقارن سبک وزن جهت صرفه‌جویی در مصرف انرژی

[1] Hyperelliptic curve
[2] Bijection
[3] Bijection
[4] Strict Avalanche Effect
[5] Algebraic Degree
[6] End-to-end

در منابع اینترنت اشیا با توان پائین با هدف دستیابی به امنیت بیشتر می‌شود [۳-۵].

الگوریتم‌های رمزنگاری سنتی برای استفاده در تلفن‌های همراه و رایانه ها بسیار مناسب هستند، در حالیکه در شبکه‌های اینترنت اشیا، نقاط انتهایی طیف شامل دستگاه‌هایی مانند سنسورها ، برچسب‌های RFID و سیستم‌های جاسازی شده است که این دستگاه‌ها بدلیل محدویت‌های حافظه و ظرفیت محاسباتی، معمولاً به سیستم عامل‌هایی نیاز دارند که از روش‌های رمزنگاری سبک وزن استفاده می‌کنند [۳]. به طور کلی، رمزنگاری سبک وزن، زیرمجموعه‌ای از روش‌های رمزنگاری است که تکنیک‌هایی را ارائه می دهد که معمولاً در دستگاه های هوشمند کم مصرف استفاده می‌شوند [۶].

تکنیک‌های رمزنگاری، بطور کلی به عنوان رمز جریانی و بلوکی [1] در نظر گرفته می‌شوند [۷]. روش‌های AES [2] [۸]، DES [3] [۱] و SMS4 [9] از گزینه‌های پرکاربرد روش‌های رمزنگاری هستند. S-Box [4] بخش اصلی در رمز بلوکی است و تأثیر مستقیمی بر سطح امنیتی رمزنگاری دارد [۱۱]. به دلیل ماهیت پویای محیط اینترنت اشیا، استفاده از رمزهای بلوکی ایستا کارایی لازم را ندارند. زیرا با توجه به اندازه بزرگ جداول S-Boxهای ایستا که در دستگاه‌های اینترنت اشیا بدلیل محدودیت‌های منابع امکان ذخیره‌سازی آنها میسر نیست، بهتر است از S-Boxها با اندازه جداول کوچک استفاده شود. از سویی دیگر،در Boxهای ایستا، بدلیل ثابت بودن رمزهای بلوکی، S-Boxهای تولید شده نسبت به S-Boxهای پویا، در برابر حملات سریع‌تر و راحت‌تر باز می‌شوند. از این رو با کوچک کردن S-Boxها و تولید S-Boxهای سبک وزن که دارای اندازه جدوال کوچکتری هستند و در عین حال پویایی آنها، می‌توان به سطح مناسبی از امنیت در دستگاه‌های با منابع محدود ذخیره‌سازی و محاسباتی دست یافت. بنابراین، پیشنهاد روشی جهت تولید S-Box پویا و سبک وزن یک نیاز حیاتی در محیط اینترنت اشیا است که در آن شبکه حسگر بی‌سیم[5] شبکه بی‌سیم بدن[6] و کارت‌های هوشمند به کار گرفته می‌شوند

بررسی‌های انجام شده بر روی مسئله S-Box پویا نشان می‌دهد که S-Box به طور گسترده‌ای بر روی محیط‌های دارای منابع حافظه و محاسباتی با ظرفیت بالا متمرکز شده است. اما هنگامی که اشیا هوشمند با عمر باتری کم، قدرت محاسبه پائین، پهنای باند و حافظه محدود که به ویژه در محیط‌های اینترنت اشیا استفاده می‌شوند، استفاده و بکار گیری S-Boxها به عنوان یک چالش مهم مطرح می‌شودکه کاملاً ناشی از توسعه فن‌آوری استفاده از دستگاه‌های مجهز به حسگرها در محیط اینترنت اشیا است [۱۳ ، ۱۴]. با توجه به محدودیت منابع حافظه و منابع محاسباتی در بستر اینترنت اشیا در دنیای واقعی، در این مقاله، بر روی رروش‌های تولید S-Box سبک وزن پویا ، جهت غلبه بر محدودیت‌های ذکر شده، تمرکز شده است.

بعنوان یکی از روش‌های مطرح در این زمینه، خم ابر بیضوی[7] توسط کوبلیتز [۱۵] برای استفاده در رمزنگاری به عنوان جایگزینی مناسب برای منحنی‌های بیضوی پیشنهاد شد. خم ابربیضوی در گروه منحنی‌های جبری است که به عنوان گسترشی از منحنی‌های بیضوی در نظر گرفته می‌شوند. همچنین، تعریف رمزنگاری خم ابربیضوی بر اساس منحنی‌هایی است که در آنها $g \geq 1$ [8] است. آنچه که استفاده از این نوع منحنی را مطلوب می‌کند آن است که خم ابربیضوی، اشیا هوشمند را قادر می‌سازد که به پهنای باند و ذخیره سازی کمتری نیاز داشته باشند [۱۶].

در این مقاله، یک روش جدید تولید S-Box پویا ارائه شده است که در آن بهبود بیشتری در امنیت روش رمزنگاری بلوک سبک وزن نسب به روش مطرح شده در SMS4 که در آن S-Box بصورت ایستا است، حاصل شده است. در این مقاله پس از بیان نحوه ساخت S-Box، به تجزیه و تحلیل آن جهت ارزیابی قدرت رمزنگاری روش پیشنهادی می‌پردازیم. ارزیابی کارایی طرح ارائه شده با انجام فرایندهای شبیه سازی با استفاده از SageMath جهت محاسبه خم ابربیضوی انجام می‌شود و سپس S-Box تولید شده تحلیل و ارزیابی می‌شود [۱۷].

به طور خلاصه، هدف این مقاله ارائه پیشنهادی یک الگوریتم جهت تولید S-Box وابسته به کلید پویا با استفاده از خم ابربیضوی است. به طور خاص ، هدف اصلی ما بهبود امنیت SMS4 از طریق ایجاد S-Box پویا و وابسته به کلید است. به‌عنوان مثال، یک S-Box موجود در SMS4 جدول ۱ نشان داده شده است [9].

جدول ۱- S-Box در SMS4 .

| x | 0 1 2 3 4 5 6 7 8 9 A B C D E F |
|---|---|
| S(x) | C 5 6 B 9 0 A D 3 E F 8 4 7 1 2 |

ساختار این مقاله بدین شرح است: در بخش ۲، مفاهیم مقدماتی روش رمزنگاری سبک وزن توضیح داده شده است. کارهای پیشین در بخش۳ مورد بحث و تجزیه تحلیل قرار گرفته است. در بخش ۴، جزئیات طرح پیشنهادی جهت تولید S-Box پویا وابسته به کلید ارائه شده است. در بخش ۵، به تجزیه و تحلیل الگوریتم پیشنهادی پرداخته شده و در بخش۶ نتیجه گیری کلی ارائه شده است.

# ۲- مفاهیم مقدماتی الگوریتم‌های رمزنگاری سبک وزن

به طور کلی الگوریتم های رمزنگاری سبک وزن بر اساس سه مفهوم اساسی بنا شده اند که عبارتند از: رمز بلوکی سبک وزن[9] ، توابع درهم ساز سبک وزن[10] و رمز جریانی سبک وزن[11].

- رمز بلوکی سبک وزن: به طور کلی، رمزنگاری بلوکی یک الگوریتم قطعی در رمزنگاری است که روی مجموعه‌ای از بیت‌ها با طول ثابت به نام بلوک‌ها با یک تغییر ثابت که از طریق یک کلید متقارن مشخص می‌شود، انجام می‌شود. همچنین، رمزهای بلوکی سبک وزن به عنوان عناصر اصلی در توسعه روش‌های مختلف رمزنگاری استفاده می‌شوند و به طور گسترده در انجام رمزنگاری داده‌های انبوه استفاده می‌شوند. به طور معمول، یک رمز بلوکی، از یک جفت تابع شامل تابع رمزنگاریری (E) و تابع رمزگشایی (D) تشکیل شده است. این توابع با ورودی‌هایی که شامل یک بلوک با اندازه n بیت و یک کلید با اندازه k بیت هستند، یک بلوک خروجی با اندازه n بیت تولید می‌کنند. تابع رمزگشایی Dبه عنوان معکوس تابع

---

1 Stream and block cipher
2 Advanced-Encryption-Standard
3 Data-Encryption-Standard
4 Substitution box
5 Wireless Body Area Network (WBAN)
6 RFID Wireless Sensor Network (RFID WSN)
7 Hyperelliptic curve
8 genus
9 Lightweight Block Ciphers (LwBC)
10 Lightweight Hash Functions (LwHF)
11 Lightweight Stream Ciphers (LwSC)

رمزنگاری E در نظر گرفته می‌شود ($E^{-1} = D$) [۱۹]. توابع رمزنگاری و رمزگشایی رمز بلوکی بوسیله معادلات ۱ و ۲ تعریف می‌شوند [۲۰].

تابع رمزنگاری:

$$E_k(P) := E(K,P) \; : \{0,1\}^k \times \{0,1\}^n \to \{0,1\}^n \qquad (۱)$$

تابع رمزگشایی:

$$E_k^{-1}(C) := D_k(C) = D(K,C) \; : \{0,1\}^k \times \{0,1\}^n \to \{0,1\}^n \qquad (۲)$$

تابع رمزنگاری (معادله ۱) ، یک کلید K به طول k و یک بیت رشته P با طول n یک رشته C با اندازه n بیت ایجاد می‌کند. P متن ساده[1] لست و C متن رمزنگاری[2] شده است. برای هر کلید K، معادله $E_k(P)$ ، باید یک نگاشت وارون بر روی $\{0,1\}^n$ باشد. در تابع رمزگشایی (معادله ۲) که معکوس E است، ورودی ها از یک کلید K و یک متن رمز دار C تشکیل شده و خروجی یک متن ساده P است، بطوریکه $\forall K: D_k(E_k(P)) = P$ .

یکی از اهداف مورد توجه در بستر اینترنت اشیا که منابع ظرفیت محاسباتی و ذخیره‌سازی محدود ی دارند، ایجاد رمزهای بلوکی سبک وزن است که به دلیل اندازه کوچکتر جداول، مناسب اینگونه بسترها هستند. بنابراین، نکات مهمی در ساخت رمزهای بلوکی سبک وزن، باید در نظر گرفته شوند که عبارتنداز:

۱) باید از بلوک‌های با اندازه‌های کوچکتر برای استفاده در رمزنگاری‌های بلوک سبک وزن استفاده شود که این موضوع اندازه متن ساده را محدود می کند.

۲) برای دستیابی به مصرف کمتر انرژی بدلیل عمر محدود باتری، باید از اندازه کلید کوچکتری استفاده شود.

۳) برای انجام مراحل محاسبه ساده‌تر در مقایسه با روش‌های رمزنگاری سنتی، باید چرخش‌های کوتاه‌تری انجام شوند. به عنوان نمونه، S-Boxها با اندازه ۴ بیت در یک S-Box سبک وزن، در مقایسه با S-Box ۸ بیتی در رمزنگاری سنتی، تولید می‌شوند.

۴) برای کاهش حافظه ومصرف کمترانرژی در تولید S-Boxها، باید روش‌های تولید کلید ساده‌تری در نظر گرفته شود زیرا یک رمزنگارز بلوک سبک وزن دارای روش‌های تولید کلیدهای ساده‌تری است که زیرکلیدها را تولید می‌کنند.

• توابع درهم ساز سبک وزن : توابع درهم ساز سبک وزن در مقایسه با تابع درهم ساز معمول که برای دستگاه‌های هوشمند اینترنت اشیا با منابع محدود، مصرف انرژی بالا را به دنبال دارند، به طور گسترده‌ای مورد استفاده قرار گرفته‌اند. اهداف اصلی طراحی توابع درهم ساز سبک وزن شامل موارد زیر است [۱۹]:

۱) اندازه کوچکتر خروجی در کاربردهایی که دارای مقاومت در برابر تصادم تابع درهم‌ساز هستند ، یک ضرورت مهم است. درکاربرد هایی که مقاومت در برابر تصادم ضروری نیست، به طور کلی از اندازه‌های متعادل استفاده می‌شود.

۲) در بستر اینترنت اشیا باید از اندازه پیام کوچکتر استفاده شود. تابع درهم ساز کلاسیک معمولاً از اندازه ۲۶۴ بیت در مقایسه با ظرفیت کمتر تابع درهم ساز سبک وزن، استفاده میکند. بنابراین ، توابع درهم ساز که برای پیام‌های کوچک استفاده می‌شوند، درکاربرد های اینترنت اشیا با ظرفیت محدود مناسب‌ترند.

• رمزنگارهای جریان سبک وزن، باید به عنوان روش‌های اصلی در محیط اینترنت اشیا با منابع محدود در نظر گرفته شوند.

در رمزهای بلوکی، یک S-Box که عنصری اصلی در روش‌های کلید متقارن است، عمل جایگزینی انجام می‌شود. در رمزهای بلوکی معمولاً ارتباط بین متن رمز و کلید ، پنهان می‌شود. به طور کلی ، یک S-Box از مجموعه ای از n بیت به عنوان ورودی استفاده می‌کند و آنها را به مجموعه ای از m بیت به عنوان خروجی تبدیل می‌کند، بطوری که ممکن است m برابر با n نباشد. یک n × m S-Box به صورت یک جدول که شامل $2^n$ کلمه m بیتی است، ساخته می‌شود [۲۱]. بطور معمول در DES، از جدول‌های ثابت استفاده می‌شود، در حالیکه در بعضی از رمزها، جداول به صورت پویا، از کلیدهایی تولیدی با روش‌هایی مانند رمزنگاری Twofish و Blowfish ساخته می‌شوند [۲۲].

## ۳- کارهای پیشین

در این بخش، به بحث و تحلیل تعدادی از مطالعات و مقالات مرتبط پیشین در رابطه با روش‌های تولید S-Box می‌پردازیم.

در [۲۳] ، نویسندگان روشی را جهت تولید S-Box پویا وابسته به کلید ارائه کردندکه با روش تولید S-Box در AES، به عنوان یک استاندارد و معیار جهت ارزیابی، مورد مقایسه قرار گرفت. روش پیشنهادی بر اساس کلمه رمز[3] تولید شده از کلید، طراحی شده است. همچنین ، کلید استفاده شده برای رمزنگاری با اندازه ۶۴ بیت در نظر گرفته شده است. یک کلمه مانند (C8C7C6C5C4C3C2C1) در زمان اجرا بر اساس فاصله و وزن همینگ کلید تولید می‌شود. این تابع شامل عملیات تغییر سطرها و ستون ها و همچنین مبادله و تبادل عناصر است. نقطه ضعف بارز روش پیشنهادی این است که به دلیل طراحی مشابه AES جهت استفاده در دستگاه‌های با منبع محدود، مناسب نیست.

در [۲۴] ، نویسندگان روشی را برای تولید S-Box وابسته به کلید بر اساس استراتژی AES، از طریق استفاده از چرخش[4] در S-Box ارائه کردند. فرآیندهای رمزنگاری و رمزگشایی در این روش مانند AES استاندادر است ، با این وجود روش موجود در AES استاندارد از چهار مرحله تشکیل شده است در حالیکه روش جدید شامل پنج مرحله است که مرحله اضافه شده در این مقاله شامل چرخش S-Box است و مقدار چرخش حاصل با کل چرخش کلید در رابطه است. محدودیت اصلی این طرح همانند محدودیت در مقاله [۲۳] است.

در [۲۵] ، یک روش جدید تولید S-Box از AES ، با استفاده از روش نگاشت متغیر ارائه شده است. روش پیشنهادی یک روش مبتنی بر AES است که داده کلید برای تولید فاکتوری استفاده می‌شود. در این الگوریتم، از مجموعه داده‌های زیر کلید وابسته به کلید اصلی، جهت نگاشت مجدد و جایگزینی S-Box به یک موقعیت تصادفی استفاده می‌شود. این طرح از چهار چند جمله‌ای غیرقابل کاهش با درجه هشت استفاده می‌کند. تغییرات لازم برای نگاشت معکوس S-Box، رابطه غیر خطی

[1] Plain text
[2] Cipher text
[3] code-word
[4] rotation

بین S-Box و عکس آن را حفظ میکند. نقطه ضعف این روش همانند ضعف روش ارائه شده در مقاله [۲۳] است.

در [۲۶] ، نویسندگان الگوریتمی جدید برای تولید S-Box پویا از طریق یک فرآیند دو مرحله‌ای معرفی کردند. در مرحله اول، از روش AES برای تولید S-Box استفاده می‌شود و در مرحله دوم، سطرها و ستون ها تبادل می‌شوند. بعلاوه، برای تولید S-Box پویا ، از کلید وابسته استفاده می‌شود. با توجه به روش AES بکار رفته جهت تولید S-Box ، روش پیشنهادی در این مقاله برای دستگاه‌های با منابع محدود مناسب نیست.

همچنین در [۲۷]، الگوریتمی برای تولید S-Box پویا پیشنهاد شده است که علاوه بر استفاده از سه تابع بازخورد خطی مختلف، از سه ثبات جابجایی[1] با بازخورد خطی استفاده می‌کند.. در خروجی این ثبات‌ها، یک عمل XOR انجام می‌شود و سپس خروجی پویا به ۱۲۸ بیت بلوک مجزا تقسیم می‌شود که از هر بلوک جهت تولید S-Box استفاده می‌شود.

# ۴- روش پیشنهادی جهت تولید S-Box

در این بخش ، ابتدا زمینه ریاضی روش تولید S-Box وابسته به کلید به طور خلاصه شرح داده می‌شود و سپس روش پیشنهادی مبتنی بر خم ابربیضوی ارائه می‌شود.

## ۴-۱- مفاهیم ریاضی

جهت روشن شدن مفاهیم اصلی کاربردی در این مقاله ، توضیح مختصری از زمینه ریاضیات مورد نیاز برای توسعه روش تولید S-Box وابسته به کلید ارائه می‌شود. روش پیشنهادی بر اساس تعاریف ارائه شده ۱ و ۲، و ایده خم ابربیضوی، متکی است.

تعریف ۱: یک خم ابر بیضوی C از نوع d بر روی میدان کامل K با مشخصه عدد اول $p$ ، بر اساس معادله ۳ تعریف می‌شود:

$$C: Y^2 + H(x)y = F(x) \qquad (۳)$$

در معادله ۳، H(x) یک چند جمله‌ای از درجه d و F(x) یک چند جمله‌ای از درجه 2d + 1 است [۲۸].

مثال ۱: اگر p =۱۱ باشد، معادله $y^2 = x^5 + 2x^2 + x + 3$ روی میدان کامل $K$ ، یک خم ابربیضوی از درجه ۲ می‌دهد.

با داشتن یک نقطه مانند P از مرتبه n در یک خم ابربیضوی C بر روی میدان متناهی $K_a$ و یک نقطه Q روی C، می‌توانیم یک عدد صحیح $m$ پیدا کنیم بطوریکه $0 \leq m \leq n-1$ و $Q = m.P$ باشد، به گونه‌ای که Q از ضرب اسکالر m و P بدست می‌آید. با دانستن مقادیر P و Q یافتن مقدار $m$ غیرممکن است. به این مشکل، مسأله لگاریتم گسسته خم ابربیضوی (HCDLP)[2] گفته می‌شود. ایده اصلی در این مقاله از همین چالش به دست آمده است.

فرآیندهای رمزنگاری و رمزگشایی از مجموعه متناهی نقاط بر روی خم ابربیضوی، روی میدان کامل K ، استفاده می‌کنند. از معادله ۳ برای به دست آوردن نقاط مانند P روی منحنی C استفاده می‌شود.

تعریف ۲: مقسوم علیه $D$ از حاصل جمع نقاط در C، بر اساس معادله ۴ به دست می‌آید:

$$D = \sum_{P\in C} m_p . P \quad , \; m_p \in Z \qquad (۴)$$

در این روش، یک عدد تصادفی به عنوان یک کلید خصوصی در نظر گرفته می‌شود و کلید عمومی Q، با ضرب کلید خصوصی در نظر گرفته شده با یک نقطه مانند P روی منحنی C بدست می‌آید. امنیت تابع رمزنگاری مبتنی بر خم ابربیضوی به سطح پیچیدگی HCDLP بستگی دارد.

کارایی تابع رمزنگاری مبتنی بر خم ابربیضوی، بر اساس محاسبه کارآمد ضرب اسکالر $Q = m.P$ است. تابع رمزنگاری مبتنی بر خم ابربیضوی از کلید با اندازه کوچک استفاده می کند که این خود باعث می‌شود در این روش، سطح امنیت مشابه الگوریتم‌های دیگر مانند RSA ارائه شود.

## ۴-۲- الگوریتم پیشنهادی

فرض کنید H یک خم ابربیضوی است که در یک میدان متناهی $K_a$ با مشخصه $a > 0$ در نظر گرفته می‌شود. فرض کنید که $D_m$ یک مقسوم علیه مرتبه n است. با توجه به $D_n$، مسأله HCDLP، شامل دستیابی به یک عدد صحیح δ است، بطوریکه $0 \leq \delta \leq n-1$ ، به گونه‌ای که $D_n = \delta . D_m$ [۲۹]. روش پیشنهادی در الگوریتم ۱ خلاصه شده است.

الگوریتم ۱: الگوریتم پیشنهادی

**Require:**
K: *Perfect field with characteristic prime number of p ;*
C: *Equation of Hyperelliptic curve 'C' over perfect field K;*
P: *Point P on the curve C;*
δ : *4- bit Key.*

**Ensure:**
Sbi : *Dynamic S-Box i.*

**Begin**
1: Calculate $D_m = \sum_{p\in c} m_p . P$
2: Calculate $D_n = \delta . D_m$
3: Calculate $D_m + D_n = \sum_{p\in c} m_p . P + \sum_{p\in c} n_p . P = \sum_{p\in c} (m_p + n_p) . P = (x_p, y_p)$
4. Obtain $Q = x_p \oplus y_p$
4. Represent Q in hexadecimal form
5. Extract n unique bits from Q into 16 hex digits of 4-bits
6. The outcome is $Sb_1$ which is 64 bits long
6. To obtain the next N dynamic S-Boxes
    For i=1 to N
        $Sb_i$= Shift S-Box $Sb_1$, 4-bits towards left circularly.
**End.**

مثال ۲: اگر در میدان کامل $K$ ، مقدار مشخصه برابر با $p = 10^{34} + 1233$ باشد، در معادله $y^2 = x^5 + 2x^2 + x + 3$، دو نقطه $P_1$ و $P_2$ را برای محاسبه $D_m$ انتخاب می‌کنیم. علاوه بر این، یک کلید برای $D_n$ تولید می‌کنیم و S-Box تولید می‌شود:

[1] linear feedback shift register

[2] Hyperelliptic Curve-Discrete-Logarithm-Problem

```
P1 = [280269558793776638909191002790764 0,
1774270760270397702615725439 21716]
P2 = [10000000000000000000000000000001231,
1312613312958640035216487254585311]
Key = 23534739862384236842
S-Box = [1, 0, 1, 0, 1, 1, 0, 0, 1, 0, 1, 1, 0, 1, 0, 0]
```

در روش پیشنهادی، برای به دست آوردن S-Boxهای بعدی، در داخل یک حلقه for، از عمل شیفت چرخشی ۴ بیت به سمت چپ استفاده می‌کنیم.

# ۵- ارزیابی کارایی روش پیشنهادی

در این بخش ، اثربخشی روش پیشنهادی از طریق چهار معیار شامل دو سویی بودن، اثر فرو پاشی بهمنی، غیرخطی بودن و درجه جبری ارزیابی می شود. سپس نتایج به دست آمده با توجه به معیارهای ذکر شده ارائه می‌شود.

## ۵-۱- معیارهای ارزیابی

عملکرد الگوریتم پیشنهادی S-Box با استفاده از معیارهای زیر ارزیابی می شود [۸]:

الف) دو سوئی بودن تابع: تابع دوسوئی $f: A \to B$ ، یک نگاشت یک به یک و پوشا از اعضای مجموعه A به اعضای مجموعه B است.

ب) اثر فروپاشی بهمنی: یک تابع، اثر فروپاشی بهمنی را ارضا می‌کند زمانی که با تغییر در یک بیت در ورودی، باعث ایجاد تغییر در بیت خروجی با احتمال ۵۰٪ شود.

ج) غیر خطی بودن: S-Box تولید شده باید ماهیتی کاملاً غیرخطی داشته باشد که بتواند فرایند تجزیه و تحلیل رمزنگاری را کاملاً دشوار کند. فرض کنید $ax + \partial$ مجموعه ای از تمام توابع ترکیبی باشد که در آن $a \in F_2^n$ و $\partial \in F_2$ . همچنین، فرض کنید $b.F = b_1 f_1 + \cdots + b_n f_n$ یک ترکیب خطی از توابع بولین $f_i$ از $F$ است که در آن $b = (b_1, \ldots b_1) \in F_2^n$, غیر صفر است. غیرخطی بودن (NL) برای یک S-Box به شکل معادله ۵ تعریف می‌شود [۳۰]:

$$NL(F) = \min\ d_H\,(b.F(x), a.x + \partial) \qquad (۵)$$

غیرخطی بودن یک S-Box با اندازه $n.m$ ، حداقل فاصله همینگ بین مجموعه تمام ترکیبات خطی غیر ثابت از توابع مولفه $F$ و مجموعه تمام توابع ترکیبی روی $F_2^n$ است.

د) درجه جبری: درجه یک تابع بولین درجه‌ای از بزرگترین تابع تک جمله‌ای در فرم نرمال جبری آن است. S-Box باید درجه جبری بالایی داشته باشد. S-Box با درجه پایین، به حملات حساس است.

چند دستور Sage جهت محاسبه درجه جبری SMS4 به این قرار است:

```
Sage= from sage.crypto.sbox import SBox
Sage=SBox [0xc,0x5,0x6,0xb,0x9,0x0,0xa,0xd,0x3,0xe,0xf,
0x8,0x4,0x7,0x1,0x2])
sage: S.min_degree()
sage: S.max_degree()
```

در این مقاله، مجموعه‌ای از آزمایشات با استفاده از SageMath جهت ارزیابی عملکرد الگوریتم پیشنهادی S-Box انجام شده است .طرح پیشنهادی با الگوریتم SMS4 در [32] مقایسه شده است. محاسبات انجام شده جهت ارزیابی روش پیشنهادی، با دو کلید مختلف انجام شده است و نتایج در جدول ۲ نشان داده شده است.

جدول 2. مقایسه نتایج ارزیابی S-Box پیشنهادی و SMS4

| الگوریتم | غیر خطی بودن | حداقل درجه جبری | حداکثر درجه جبری |
|---|---|---|---|
| SMS4 | ۴ | ۲ | ۳ |
| S-Box پیشنهادی | ۴ | ۴ | ۴ |

با توجه به نتایج به دست آمده، چهار معیار ارزیابی دو سویی بودن، اثر فرو پاشی بهمنی، غیرخطی بودن و درجه جبری الگوریتم پیشنهادی بررسی می‌شود.

الف) دو سوئی بودن الگوریتم پیشنهادی:

در S-Box وابسته به کلید پویای پیشنهادی، نگاشت یک به یک و پوشا، از بردارهای ورودی به بردارهای خروجی وجود دارد، زیرا بردارهای ورودی و بردارهای خروجی ایزومورف هستند. بنابراین معیار دو سوئی بودن برای S-Box وابسته به کلید پویا پیشنهادی برآورده شده است.

ب) اثر فروپاشی بهمنی الگوریتم پیشنهادی:

در الگوریتم پیشنهادی، تغییرات جزئی در بردار ورودی منجر به تغییر چشمگیر در بردار خروجی می‌شود. S-Box برای کلیدهای ۲۳۵۳۴۷۳۹۸۶۲۳۸۴۲۳۶۸۴۳ و ۲۳۵۳۴۷۳۹۸۶۲۳۸۴۲۳۶۸۴۲ (باتغییر یک بیت) [۱ ، ۱ ، ۱ ، ۰ ، ۰ ، ۰ ، ۱ ، ۰ ، ۰ ، ۰ ، ۱ ، ۱ ، ۱ ، ۰ ، ۰ ، ۰ ، ۰] ، [۱ ، ۰ ، ۱، ۰ ، ۱ ، ۱ ، ۰ ، ۰ ، ۱ ، ۰ ، ۱ ، ۱ ، ۰ ، ۱ ، ۰ ، ۰] است و از این رو بطور قابل توجهی S-Boxهای مختلفی تولید می‌شوند. بنابراین فروپاشی بهمنی در کلید پویا وابسته به کلید S-Box پیشنهادی وجود دارد و کامل است.

ج) غیر خطی بودن الگوریتم پیشنهادی:

از آنجا که کلیدهای پویای وابسته به S-Boxها کاملاً تصادفی از کلید تولید می‌شوند، هر S-Box دارای خصوصیت غیرخطی نسبتاً بالایی است و احتمال کامل بودن آن بالاست. غیرخطی بودن الگوریتم پیشنهادی در جدول ۲ نشان داده شده است. در [31]، نویسندگان اظهار داشتند که غیرخطی بودن باید به بهترین غیرخطی شناخته شده نزدیک باشد (به عنوان مثال، مانند NL=۴ که توسط SMS4 S-Box بدست آمده). بنابراین، در این مقاله، NL > ۳ را برای S-Box بدست آمده که از نظر رمزنگاری، درطبقه قوی قرار می‌گیرد.

د) درجه جبری الگوریتم پیشنهادی:

نتایج بدست آمده نشان می‌دهد که در الگوریتم پیشنهادی، مقدار درجه جبری برابر با ۴ است که عملکرد بهتری نسبت به الگوریتم SMS4 از خود نشان می‌دهد.

# ۶- نتیجه‌گیری

بخش اصلی هر روش رمزنگاری کلید متقارن، S-Box است که با چالش‌هایی مواجه است. در اکثر الگوریتم‌های رمزنگاری، S-Box ثابت است. چالش این مقاله، چگونگی تضمین کارایی S-Box تولید شده در دستگاه‌های پویای اینترنت اشیا با منابع محدود است. در طرح پیشنهادی، از خم ابر بیضوی برای تولید S-Box استفاده

شد و نیز ارزیابی الگوریتم پیشنهادی با استفاده از ابزار Sage انجام پذیرفت. با توجه به ارزیابی‌های انجام شده این نتیجه حاصل شد که طرح پیشنهادی برای تولید S-Box وابسته به کلید پویا، کلیه معیارهای یک S-Box کارآمدتر را نسبت به SMS4 را برآورده می‌کند.

## ۷- مراجع